\documentclass[twocolumn,showpacs]{revtex4}
\usepackage{graphicx}
\usepackage{dcolumn}
\input epsf
\def\be{\begin{equation}}
\def\ee{\end{equation}}
\def\bea{\begin{eqnarray}}
\def\eea{\end{eqnarray}}

\def\Re{\mathop{\rm Re}\nolimits}
\def\Im{\mathop{\rm Im}\nolimits}

 \newcommand{\ft}[2]{{\textstyle\frac{#1}{#2}}}
 \newcommand{\Ka}{K{\"a}hler}
\newsavebox{\uuunit}
\sbox{\uuunit}
    {\setlength{\unitlength}{0.825em}
     \begin{picture}(0.6,0.7)
        \thinlines
        \put(0,0){\line(1,0){0.5}}
        \put(0.15,0){\line(0,1){0.7}}
        \put(0.35,0){\line(0,1){0.8}}
       \multiput(0.3,0.8)(-0.04,-0.02){12}{\rule{0.5pt}{0.5pt}}
     \end {picture}}

\newcommand{\rmi}{{\rm i}}

\begin{document}
 \tighten

\

\preprint{SU-ITP-03/01}
\title {\Large\bf SuperCosmology }
 \author{\bf Renata Kallosh and Sergey Prokushkin }
\address{  {Department of Physics, Stanford University, Stanford, CA 94305-4060,  USA}    }

{\begin{abstract}

We present a Mathematica package for performing algebraic and numerical computations
in cosmological models based on supersymmetric theories. The programs allow for
(I) evaluation and study of the properties of a scalar potential in a large class of supergravity models 
with any number of moduli, an arbitrary superpotential, \Ka\ potential, and  D-term; 
(II) numerical solution of a system of scalar and  Friedmann equations for the flat FRW universe
with any number of scalar moduli and arbitrary moduli space metric. We are using here a simple set of first order differential equations which we derived in a Hamiltonian framework. 
Using our programs we present some new results: 
(I) a shift-symmetric  potential of the inflationary model with a mobile D3 brane in an internal space with 
stabilized volume;  
(II) a KKLT-based dark energy model with the acceleration of the universe due to the evolution of 
the axion partner of the volume modulus.  
The gzipped package can be downloaded from \url{http: //www.stanford.edu/~prok/SuperCosmology/} 
or from  \url{http: //www.stanford.edu/~rkallosh/SuperCosmology/}

\end{abstract}}
\maketitle
\section{Introduction}
The studies of the cosmological aspects of supergravity and
string theory have a long history, going back to the beginning of
the 80's.  At present, there is a new wave of interest in the
cosmological aspects of string theory. The subject is rather
complicated, partly due to
the complexity of the analytical study of these theories. For
example, to find the F-term
part of the effective potential in supergravity,
one should specify the expressions for the \Ka\  potential $K$ and
superpotential $W$. Even with the simplest  $K$
and $W$, the computation of  potential is tedious, especially if there is more than one superfield, and the resulting expression is hard to analyse.

Deriving cosmological consequences from the models based on string theory and supergravity is intricate too. For
non-canonical \Ka\ potentials (which are the rule rather than the
exception) the equations of motion acquire additional
velocity-dependent terms whose effects are not easily understood using  
intuition based
on  the simplest  scalar field models. For the case of
one field, one can always reduce the theory to the canonical form,
but  this method does not work for the description of the simultaneous
motion of several different fields, so one should really solve the
system of equations keeping the non-canonical kinetic terms throughout.

Our Mathematica-based package ``SuperCosmology'' is
intended to simplify the study of supergravity potentials and
of the cosmological models based on string theory and
supergravity. The programs we describe here were used in papers
\cite{Kallosh:2002gf}-\cite{Hsu:2003cy}, where only the final
results of computations were presented. The purpose of this paper
is to present the explanation of the programs and methods we used
in computations. Also we present some new cosmological models and
use them to demonstrate how our package works.

The SuperCosmology package consists of two parts. 
 Part~I has the
following Mathematica nb-files:\\

\noindent {\tt SuperPotential.nb}, \\
{\tt SuperPotential\_KKLT.nb}, \\
{\tt SuperPotential\_fine\_tune.nb}, \\
{\tt SuperPotential\_D3.nb}. \\

In {\tt SuperPotential.nb}, {\tt SuperPotential\_KKLT.nb},
{\tt SuperPotential\_fine\_tune.nb} one finds examples from \cite{Hsu:2003cy},
\cite{Kachru:2003aw} and  \cite{Kachru:2003sx}, respectively,  of
using our  program ``SuperPotential'' in computation of a scalar
potential. In  {\tt SuperPotential\_D3.nb}, a new example of an inflationary
potential with a mobile D3 brane is studied, which is based on the
D3/D7 inflationary model investigated before in
\cite{Herdeiro:2001zb}-\cite{Hsu:2004hi}.

\noindent In part~II,  we present the program ``FRW'' used  in \cite{Kallosh:2002gf} for the numerical solution of  the Friedmann equations for a system with any number of scalar fields with geometric kinetic terms of the form 
${1\over 2} G_{ij} (\phi, \phi^*) \partial\phi^i \partial \phi^{j}$ specified by a metric $G_{ij} (\phi)$
on the scalar manifold. Part~II has the following Mathematica nb-files:\\ 

\noindent {\tt FRW\_N2.nb}, \\
{\tt FRW\_DarkE.nb}, \\
{\tt FRW\_LateDarkE.nb}. \\

In  {\tt FRW\_N2.nb}, we show a dark energy model based on the N=2 supergravity model  \cite{Fre:2002pd}, 
\cite{Kallosh:2002wj} which was also discussed in \cite{Kallosh:2002gf}. 

New results on the dark energy model based on the KKLT model \cite{Kachru:2003aw},
are presented in {\tt FRW\_DarkE.nb} and {\tt FRW\_LateDarkE.nb}. 
The examples in {\tt FRW\_DarkE.nb}  describe the situation 
when the system has not yet reached the dS minimum  (so that the scalars are still moving and $\Omega_D$ is still increasing). In {\tt FRW\_LateDarkE.nb} we study the long term evolution including the time when the dS minimum is reached by the scalars.

\section{N=1 supergravity potentials fixing the moduli}

It is believed that an N=1 d=4 supergravity may serve as an effective theory for a more fundamental string/M theory. 
A generic supersymmetric gravity has two types of geometries: the space-time geometry 
$$ds^2=g_{\mu\nu}(x) dx^\mu dx^\nu,$$ 
and the moduli space geometry
$$ds^2= G^i{}_j(\phi) d\phi_i d \phi^j,$$
which both affect  the cosmological models.
Here we consider a class of models of N=1 supergravity with any number $n_c$ of chiral multiplets $(\phi_i\,, \phi^i=(\phi_i)^*\,, \; i=1, ..., n_c)$, one Abelian vector multiplet and a gravitational multiplet.  In application to cosmology, one primarily needs to know the total potential of the scalar fields and their kinetic terms 
(since these are typically non-canonical). In the general case of many vector multiplets, the kinetic function is a matrix $f_{\alpha \beta}(\phi)$ where $\alpha, \beta=1,..., n_v$ with $n_v$ the number of vector multiplets. 
Here we will consider a simpler picture with only one vector multiplet and one holomorphic function $f(\phi)$.

The supergravity action is defined by the functions \cite{Cremmer:1982en}
\be\label{input}
     W(\phi)\,,  \quad  K(\phi,  \phi^*)\,,  \quad    \mbox{and} \quad f(\phi) \,. 
\ee     
(We do not include constant FI terms here, however, we will have a field-dependent D-term potential.) 
The bosonic part of the action is (we use  units in which $M_P=1$):
\begin{eqnarray}
e^{-1}{\cal L}_{\rm bos}&=&-\ft12 R -g_i{}^j(\hat{\partial }_\mu
\phi ^i)(\hat{\partial }^\mu \phi _j) -V\nonumber\\ [2mm]
 &-&\ft14(\Re f )F_{\mu \nu } F^{\mu \nu  }
 +\ft 14\rmi(\Im f)
 F_{\mu \nu } \tilde F^{\mu \nu  }\,.
 \label{bosonic}
\end{eqnarray}
The potential consists of an $F$-term and a $D$-term:
\begin{eqnarray}
     V&=&V_F+V_D\,,  \label{Vtotal}\\
     \nonumber\\
     V_F &=&{\rm e}^K \left[ ({\cal D}^iW)(g^{-1})
      {}_i{}^j({\cal D}_jW^*) -3 WW^*\right]\,, \label{VF}\\
      \nonumber\\
     V_D&=&\ft12\left.(\Re f) D D\right|_{bos}=
     \ft{1}{2}(\Re f)^{-1 }  {\cal P} {\cal P}\,,   \label{VD}
\end{eqnarray}
with
\begin{eqnarray}
{\cal D}^iW&=& \partial ^i W +(\partial ^i K) W\,, \\
\nonumber\\
 {\cal P}(\phi,\phi^*)  &=& \rmi\,\left[ \delta \phi_i
\partial^i K(\phi,\phi^*) )\right]=
\nonumber \\
 &=& \rmi\,\left[ -\delta \phi^i
\partial_i K(\phi,\phi^*) )\right] \label{Pnew} \, ,
\end{eqnarray}
where $ (\delta \phi_i, \delta \phi^i)$ defines the $U(1)$ gauge transformations of chiral superfields. The covariant derivative of $\phi_i$ is
\begin{equation}
  \hat{\partial }_\mu \phi_i= \partial _\mu \phi_i -W_\mu  \delta \phi_i\,.
\label{hatdz}
\end{equation}

Our package ``SuperCosmology'' is intended to help with studying models in the class presented above.  
In particular, it allows one to find the potential  (\ref{Vtotal})-(\ref{VD}) for a given input (\ref{input}) plus the gauge 
transformations of the superfields $\delta \phi_i$. We will explain now how to use part~I of our package by 
considering the example given in {\tt SuperPotential.nb}. 

In the file {\tt SuperPotential.nb} we use the scalar potential of the D3/D7 inflationary model with a light D7 brane
moving towards a heavy stack of D3 branes in an internal space with a stabilized volume \cite{Hsu:2003cy}.
We give as  input a list of complex fields $\rho, S, \Phi_+, \Phi_-$.
These are components of the holomorphic field $\phi_i$.  
(In the files, we use barred symbols for the anti-holomorphic fields, 
e. g. $\bar \rho$ is the complex conjugate field to $\rho$.) The real parts of the fields $\rho$ and $S$ correspond
to the volume of the internal space and the inflaton, respectively.
The next piece of input is the holomorphic superpotential 
$W(\phi)$, the \Ka\  potential $K(\phi,  \phi^*)$, and the $U(1)$ gauge transformations of the scalar fields. 
The program calculates  all important structures, including the $\sigma$-model metric $G_i{}^j(\phi,  \phi^*)$ defining the kinetic terms of 
the chiral superfields. It also  presents a matrix form of it and an expression for all covariant derivatives of the superpotential  $D_i\, W$. 
This simplifies the search for the supersymmetric configurations with constant scalar fields, since we can now easily
solve the equation  $D_i\, W=0$ in all directions $\phi_i$. Next, the program calculates the F-term and the D-term potentials as functions of all fields $(\phi,  \phi^*)$. We found it convenient for computations to switch to real 
scalar fields by splitting  the complex fields into their real and imaginary parts, 
$\rho=\sigma + i \alpha$, so that the total potential can be given as a function of $2n_c$ real scalar fields.  Finally, the program plots various sections of the potential as  functions of just two fields at fixed values of other fields.
The physical properties of this model are described in  \cite{Hsu:2003cy}. 

We supplement {\tt SuperPotential.nb} with a few more examples. The example in {\tt SuperPotential\_KKLT.nb} shows how the stabilization of the volume was achieved in \cite{Kachru:2003aw}. The example in {\tt SuperPotential\_fine\_tune.nb} presents a detailed calculation of the fine-tuning procedure for the potential described in the Appendix F of \cite{Kachru:2003sx}. 

Our fourth example presented in {\tt SuperPotential\_D3.nb}, is based on the potential for the model of inflation with D3/D7 branes 
\cite{Herdeiro:2001zb} with volume stabilization, where the D7 brane is heavy and therefore only the D3 brane 
position modulus is turned on. Note that the \Ka\ potential in the case of a heavy stack of D3 branes and 
a light D7  \cite{Hsu:2003cy}, which is used in {\tt SuperPotential.nb}, is given by
\be
     K=-3\ln(\rho+\bar \rho ) - {(S-\bar S)^2\over 2}   \,.
\ee
The opposite limiting case with a light D3 and heavy D7 branes, considered in {\tt SuperPotential\_D3.nb}, 
is described by a \Ka\ potential 
\be
     K=-3\ln\left (\rho+\bar \rho - {(\phi+\bar \phi)^2\over 2}\right)\, ,
\ee
in agreement with \cite{Hsu:2003cy}, \cite{Firouzjahi:2003zy}-\cite{Hsu:2004hi}, and the superpotential the same 
as in the KKLT model  \cite{Kachru:2003aw},
\be
      W = W_0 + Ae^{-a\rho}  \,.
\ee
The example given in the file {\tt SuperPotential\_D3.nb} describes a situation which was not explored before and turned out to
be quite attractive. We have the same ``trench'' in the $s={(\phi-\bar \phi)\over 2}$ direction which is to be interpreted 
as the inflaton direction. 
As  shown in {\tt SuperPotential\_D3.nb}, the dependence on $\alpha={(\rho-\bar \rho)\over 2i}$ is quite simple: the potential 
includes $c\cos\alpha$  with $c$ negative, and, therefore, there is a minimum at $\alpha=0$: the same behavior
as in  \cite{Hsu:2003cy}. However, the dependence on the field $\beta= {(\phi+\bar \phi)\over 2}$ in this model 
is quite different from that in  \cite{Hsu:2003cy} . Indeed, in the model studied in  \cite{Hsu:2003cy} 
(see {\tt SuperPotential.nb}), the F-term potential has a supersymmetric extremum at $\beta=0$ which is a saddle point. 
This saddle point remains after lifting the potential by adding a D-term, and a stable configuration must have 
some  $\beta\neq 0$ after $\beta$ rolls down to the closest nearby minimum. In the model presented here
in {\tt SuperPotential\_D3.nb},  the F-term potential also has a saddle point at $\beta=0$. However, here after the D-term potential 
is added, this saddle point turns into a perfect dS minimum. Since the stabilized value of the volume
modulus $\sigma_{cr}$ depends on the values of  $\alpha$ and $\beta$ at the extremum, $\sigma_{cr}$
here is the same as in the KKLT model \cite{Kachru:2003aw}, with $\alpha_{cr}=\beta_{cr}=0$. 
This gives a good starting point for a study of stringy models of  hybrid inflation (in particular, 
type IIB string theory compactified on $K3\times {T^2\over Z_2}$  with D3/D7 branes present), based on 
the same mechanism of the volume stabilization as in the model of \cite{Kachru:2003aw}.

\section{Solving Friedmann equations in supergravity-based models of dark energy}

In a series of papers \cite{Kallosh:2002gf}-\cite{Kallosh:2003bq}, the issue of dark energy has been studied 
in the context of supergravity. In particular, in these models  we have studied the future of the universe and 
compared the predictions of some models with observations. The more recent supernova data for $Z>1$ in \cite{Riess:2004nr} were also compared with the theoretical predictions in \cite{Kallosh:2003mt}, \cite{Kallosh:2003bq}.

The action of a generic four-dimensional gauged supergravity which can be used for a dark energy hidden sector,
includes gravity coupled to $2n$ scalar fields $\phi^i$, and a potential:
\be\label{real}
     g^{-1/2} L = -{1\over 2} R +  {1\over 2}\,G_{ij}(\phi)\,  \partial_\mu \phi^i \partial_\nu \phi^j\, g^{\mu\nu} -V(\phi) \ .
\ee
Here $G_{ij}(\phi)$ is the metric on the moduli space, related to the \Ka\ potential. 
(In the cosmological context it is convenient to work with $2n$ real scalar fields, as in (\ref{real}),
rather then $n$ complex fields).

We consider here a model with dark energy represented by the scalars $\phi^i$ with the Lagrangian 
(\ref{real}). Also, we include the usual cold dark matter in the model, with the energy density $\rho_M$, 
\be\label{dust}
    \rho_M= {C\over a^3}  \,. 
\ee 
Also, we assume that the space is a flat FRW universe, $ds^2= dt^2-a(t)^2 d\vec x^2$, and that the scalar fields are
homogeneous. With these assumptions, the scalar and Friedmann equations are: 
\be\label{second}
\ddot \phi^i + 3{\dot a\over a} \dot \phi^i + \Gamma^i_{jk}\dot \phi^j \dot \phi^k  + G^{ij}  {\partial V\over \partial \phi^j} =0 \ ,
\ee
\be
  {\ddot a\over a}= {-\rho_M + 2 V - 4E_{kin}\over 6}\ .
\label{F1}\ee
Here $\Gamma^i_{jk}(\phi)$ are the Christoffel symbols in the moduli space defined by the metric $G_{ij}(\phi)$,
and $E_{kin}$ is the kinetic part of the dark energy,
\begin{equation}
E_{kin} =   {1\over  2} G_{ij}(\phi)  \dot \phi^i \dot \phi^j  \,.
\end{equation}

\subsection{First order Friedmann equations}

The Friedmann equations for homogeneous scalar fields take a particularly simple form in the Hamiltonian
(first order) formulation, when the metric is a spatially flat FRW. For numerical calculations the first order equations 
are also much more suitable so we use them in our computations.

With the Lagrangian (\ref{real}) the canonical momenta are 
\be\label{} 
  P_i = a^3(t) G_{ij}(\phi)\dot \phi^i  \,,
\ee 
and the Hamiltonian is
\be
{\cal H}(P, \phi, t) = {1\over 2 a^3(t)} G^{ij}(\phi) P_i P_j    + a^3(t) V(\phi)\ .
\label{Ham} \ee 
The equations of motion have the  canonical form,
\bea
\dot \phi^i  = {\partial {\cal H}\over \partial P_i} \,,\quad 
 \dot P_i = -{\partial {\cal H}\over \partial \phi^i}   \,. 
\eea
Using the Hamiltonian (\ref{Ham}) and adding the equation for the scale factor, we get a system of the first order ODEs:
\bea
\dot \phi^i  &=&  {1\over  a^3}  G^{ij}(\phi)  P_j \label{F2} \\
\nonumber\\
 \dot P_i &=&   -  {1\over 2 a^3} {\partial G^{kl}\over \partial \phi^i} P_k P_l    -a^3 {\partial  V(\phi)\over \partial \phi^i}\label{F3}\\
\nonumber\\
\dot a &= & a\, H\label{F4}
\\
\nonumber\\
{\dot H} &= & {-\rho_M + 2 V - 4E_{kin}\over 6}- H^2
\label{F5}\eea
Here the kinetic energy $E_{kin}$ is given by 
\be
E_{kin}= {1\over  2 a^6}G^{ij}(\phi)  P_i P_j    =  {1\over  2 } G_{ij}(\phi)  \dot \phi^i \dot \phi^j  
\label{kin}
\ee
and $\rho_M$ is given in (\ref{dust}).

In the simplest case of one dark energy scalar field with a minimal kinetic term of the form $(\dot\phi)^2/2$ 
the system of equations (\ref{F2}), (\ref{F3})  reduces to a familiar form,
\bea
\dot \phi  &=&  {1\over  a^3}    P  \\
\nonumber\\
 \dot P &=&    -a^3 V' \qquad \Rightarrow \qquad \ddot \phi + 3{\dot a\over a} \dot \phi +   V' =0
\label{simple}\eea
The system (\ref{F2})-(\ref{F5}) has to be supplemented by the initial conditions for $\phi^i$, $P_i$, $a$, and $H$. 
We always set  $a(t=0)=1$, and use the first integral of motion to determine 
 $H(t=0)= \sqrt{{\rho_M + V(\phi) + (\dot\phi)^2/2\over 3}}\left.\right|_{t=0}$.

The energy density $\rho_{{}_D}$ and the pressure $p_{{}_D}$ for  scalar dark energy are given by
\bea
       \rho_{{}_D}&=& E_{kin}+ V = {1\over  2 } G_{ij}(\phi) \dot \phi^i \dot \phi^j  + V(\phi) \ ,\\
       p_{{}_D}&=& E_{kin}- V = {1\over  2 } G_{ij}(\phi) \dot \phi^i \dot \phi^j  - V(\phi) \ .
 \eea
 The total energy also includes  the energy of matter,
 \be
       \rho_{{}_T}= \rho_M+ E_{kin}+ V \ .
 \ee
The $\Omega$ parameter for dark energy (matter) is given by the ratio of the dark energy (matter) to the total energy,
\be
\Omega_{D}= {\rho_{{}_D}\over \rho_{{}_T}} \ , \qquad \Omega_{M}= {\rho_{{}_M}\over \rho_{{}_T} }\ .
\ee
Another important characteristic of the dark energy is its pressure-to-energy ratio defining the dark energy 
equation of state $p_{{}_D}= w_{{}_D} \rho_{{}_D} $,
\be
w_{{}_D}= {p_{{}_D} \over \rho_{{}_D}}= {E_{kin}- V\over E_{kin}+ V} \ .
\ee

The file {\tt FRW\_N2.nb} shows how we solve the Friedmann equations (\ref{F2})-(\ref{F5}) for  the $N=2$  supergravity model  
\cite{Fre:2002pd} as discussed in the context of dark energy in \cite{Kallosh:2002gf}. Solutions of these equations with 
various initial conditions for the scalar fields, have an attractor behavior: the scalars eventually reach their attractor values 
defined by a minimum of the potential, and the universe asymptotically becomes  
de~Sitter. 
Note that this takes place  at $t_{\rm final}\approx 3$. However, at this value of $t_{\rm final}$, when all scalars are at their critical 
values, $w=-1$ and $\Omega_D$ is greater than today's observed value $\Omega_D\sim 0.72$. 
Nevertheless, one may still use this model to describe the dark energy of the present universe, assuming that 
the scalars have not reached the critical point yet (they are still evolving towards it). 
As we see from the plots, at $t_{\rm final}< 0.8$ we have $\Omega_D\sim 0.72$, and at this time the equation 
of state function $w(t)$ takes various values, depending on the initial conditions, which are mostly different from $-1$ but not far from it. 

\subsection{A KKLT-based toy model of dark energy}

The  KKLT model \cite{Kachru:2003aw}  is given by:
\bea
&&{g}^{-1/2} L = -{1\over 2}R +{3 \partial \rho \partial \bar \rho\over (\rho+\bar \rho)^2} -V(\rho, \bar \rho)
\eea
With $\rho= \sigma +i\alpha$ the kinetic term for scalars and the potential $V(\sigma, \alpha)$ are:
$${3\over 4\sigma^2}[(\partial \sigma)^2 +(\partial \alpha)^2]$$
$$ V= {aA e^{-2a\sigma}\left (A(3+a\sigma)+3 e^{a\sigma}W_0 \cos[a\alpha]\right)\over 6\sigma^2} + {D\over \sigma^3}$$
We study this model in {\tt FRW\_DarkE.nb}. The model has a number of interesting features. In  some cases, when one starts close to the ridge, the volume first tends to increase, however, as soon as the axion starts moving, the volume comes back and stops there. For a long time it does not change, whereas the axion tends to move quickly towards its minimum and oscillates around it. In practically all cases,  $\Omega_D$ grows during the axion motion and reaches the value 0.72 at $t\approx 1$. During the same period of time $w(t)$ remains  close to -1, however it deviates from this value and takes various shapes, depending on initial conditions. We also plot the acceleration parameter, $q(t)=-{\ddot a\over a^2H^2}$.

One has to keep in mind, that any realistic dark energy model has to accommodate  today's value of the dark energy, which means that 
the minimum of the potential has to be close to $10^{-120}$. 

\begin{figure}
\centering\leavevmode\epsfysize=6.5 cm \epsfbox{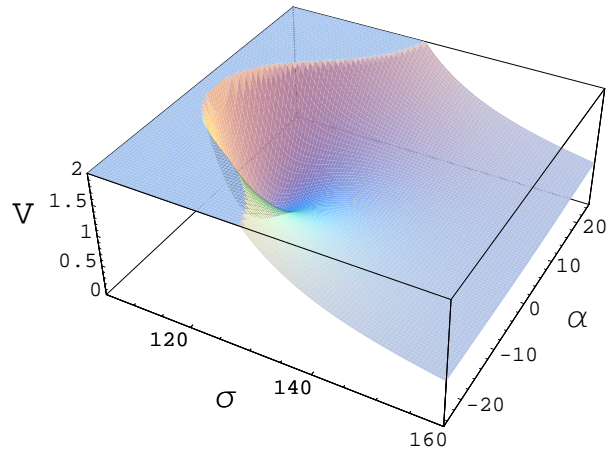}

\

\

\caption{Potential of the KKLT model depending on the volume $\sigma$ and the axion $\alpha$} \label{fig:Fig1}
\end{figure}

\begin{figure}
\centering\leavevmode\epsfysize=8.5 cm \epsfbox{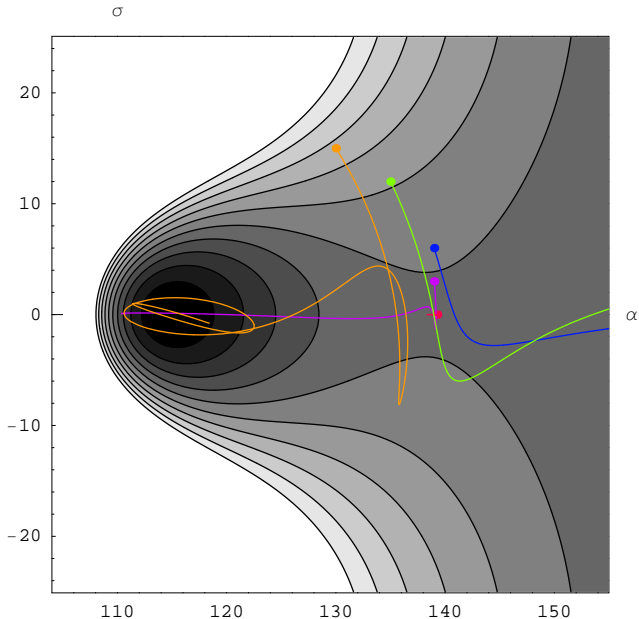}

\

\

\caption{Contour plot in the $(\sigma, \alpha)$ plane} \label{fig:Fig2}
\end{figure}

\begin{figure}
\centering\leavevmode\epsfysize=4 cm \epsfbox{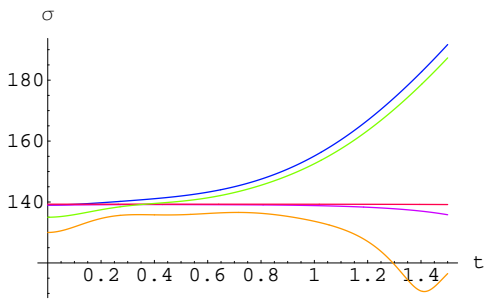}

\

\

\caption{Volume as a function of time} \label{fig:Fig3}
\end{figure}

 \begin{figure}
\centering\leavevmode\epsfysize=4 cm \epsfbox{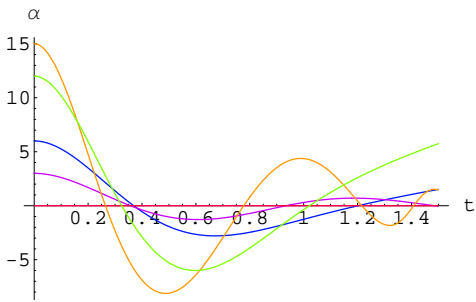}

\

\

\caption{Axion as a function of time} \label{fig:Fig4}
\end{figure}

 \begin{figure}
\centering\leavevmode\epsfysize=4 cm \epsfbox{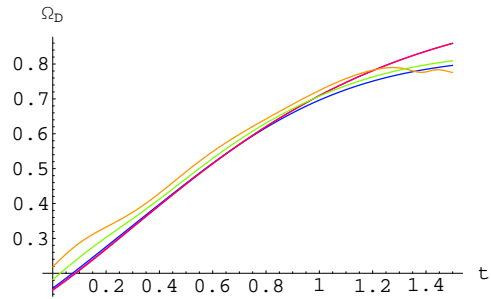}

\

\

\caption{$\Omega_D$ growth towards $\sim 0.72$} \label{fig:Fig5}
\end{figure}

  \begin{figure}
\centering\leavevmode\epsfysize=4 cm \epsfbox{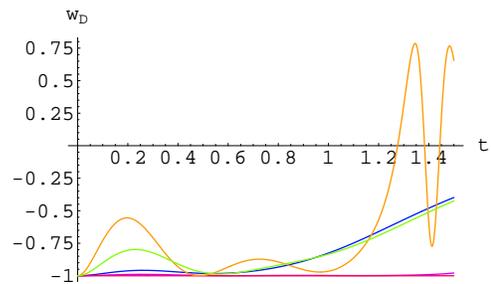}

\

\

\caption{Equation of state function $w(t)$} \label{fig:Fig6}
\end{figure}
It is also interesting to study  the late time behavior of the system. In  {\tt FRW\_LateDarkE.nb},  we have solved the same equations with 
the same initial conditions up to the time $t\approx 30$.  In most cases the system ends up at the dS minimum at $\sigma\sim 116$ and $\alpha=0$,
with $w(t)=-1$. In   other cases, both the volume and the axion have a runaway behavior, and the internal7 space de-compactifies.

\section{Discussion}
The SuperCosmology Mathematica package presented in this paper proved to be useful in numerous applications, both in our previous work as well as in the new models described in this paper.  The interesting features of the new models are due to the special choice of potentials allowed in supersymmetric theories,  and  the non-canonical geometric kinetic terms.

One of the new  models studied here is the KKLT-based model of 
dark energy. The kinetic term of the model has an $SL(2,R)$-symmetry typical for string theory and supergravity (see \cite{Horne:1994mi} for earlier studies of cosmology with $SL(2,R)$-symmetry). One can find a change of variables which will bring one of the scalars to the canonical form, $\rho= \sigma +i \alpha$, $\sigma=e^{\sqrt{2/3} \phi}$, however, the other one  cannot be canonical: 
$$L_{ kin}= 3{ \partial \rho \partial \bar \rho\over (\rho+\bar \rho)^2}={1\over 2} [(\partial\phi)^2+ {3\over 2} e^{-2\sqrt{2/3} \phi} (\partial\alpha)^2]$$
The KKLT non-perturbative potential depends on the volume modulus $\sigma$ and on the axion $\alpha$ and has a complicated profile. We have shown the contour plot of this potential as well as some trajectories of scalar fields in Fig. 2.

Consider, for example, the evolution of the model from the point $\sigma=130$, $\alpha=15$. 
This initial point corresponds to the top left (orange) dot on the Fig. \ref{fig:Fig2}. The solution obtained numerically shows that the volume modulus $\sigma$ after 
some initial  increase  stops and waits, while the axion evolves towards $\alpha = 0$ and oscillates around it. 
Then $\sigma$  moves back and eventually 
both fields get trapped at the minimum of the potential. The unusual behavior of the fields is explained by an interplay between  the potential and  non-canonical
 kinetic terms. From a more general perspective: in the second order equation for the scalars (\ref{second})
there is an extra term $\Gamma^i_{jk}\dot \phi^j \dot \phi^k$ 
in addition to the standard friction due to the Hubble parameter and also the contribution of the potential depends on the metric,  $G^{ij} {\partial V\over \partial \phi^j}$. That makes it hard to guess,  before a numerical solution of equations is found, why some trajectories end up at the minimum of the potential whereas some other trajectories lead to the volume de-compactification. 

Our package ``SuperCosmology'' is intended to aid in the study of models related  to string theory as shown in our examples. We hope it will prove 
to be useful for further investigations of the interface between string theory, supergravity and cosmology.

\subsection*{Acknowledgments}
It is a pleasure to thank  T.~Banks, C.~Burgess, K.~Dasgupta, M.~Dine, A.~Frolov, J.~Hsu, S.~Kachru, L.~McAllister, S.~Shenker, 
E.~Silverstein, S.~Trivedi, and F.~Quevedo for useful discussions which stimulated us to polish our code and make it public.  We are particularly grateful to  A.~Linde and M.~Shmakova for the collaboration on the ``Supergravity, dark energy and the fate of the universe'' project \cite{Kallosh:2002gf} where we have started using numerical methods extensively, and to A.~Van~Proeyen who shared with us a code for computing F-term potentials in supergravity 
with one chiral superfield, which was used in \cite{Kallosh:2000ve}. This work is supported 
by NSF grant PHY-0244728.

\end{document}